\documentclass[12pt]{JHEP3}

\usepackage{graphicx, cite}
\usepackage{amsmath,amssymb,amsthm,amscd}
\def\be{\begin{equation}}
\def\ee{\end{equation}}
\def\ba{\begin{array}{l}}
\def\ea{\end{array}}
\def\d{\dagger}
\def\beg{\begin{equation}}
\def\bea{\begin{eqnarray}}
\def\eea{\end{eqnarray}}

\def\nn{\nonumber}

\def\tr{\textrm{tr}}

\def\vev#1{\langle #1 \rangle}

\newcommand{\del}{\partial}

\title{Plasma balls/kinks as solitons of large $N$ confining gauge 
theories}

\author{  
Pallab Basu \\ 
\email{pallab@theory.tifr.res.in} 
}
\author{
Bobby Ezhuthachan \\
\email{bobby@theory.tifr.res.in}
}
\author{
Spenta R. Wadia \\ 
\email{wadia@theory.tifr.res.in} \\
}
\author{ Tata Institute of Fundamental Research, \\
Homi Bhaba Road, Mumbai-400005, India 
}
\preprint{TIFR/TH/06-35}

\abstract{
We discuss finite regions of the deconfining phase of a confining gauge theory (plasma
balls/kinks) as solitons of the large $N$, long wavelength, effective Lagrangian of the thermal
gauge theory expressed in terms of suitable order parameters. We consider a class of confining gauge theories whose effective Lagrangian turns out to be a generic 1 dim. unitary matrix model. The dynamics of this matrix model can be studied by an exact mapping to a non-relativistic many fermion problem on a circle. We present an approximate solution to the equations of motion which corresponds to the motion (in Euclidean time) of the Fermi surface interpolating between the phase where the fermions are uniformly distributed on the circle (confinement phase) and the phase where the fermion distribution has a gap on the circle (deconfinement phase). We later self-consistently verify that the approximation is a good one. We discuss some properties and implications of the solution including the surface tension which turns out to be positive. As a by product of our investigation we point out the problem of obtaining time dependent solutions in the collective field theory formalism due to generic shock formation.  
}

\begin{document}

\section{Introduction}
The study of black holes using the AdS/CFT was initiated by Witten's observations that black hole spacetimes in $AdS_5$ are holographic duals of the deconfinement phase of the four dimensional SU(N) gauge theory on the 
boundary of $AdS_5$ \cite{Witten:1998zw}. Various subsequent studies explored this connection to discuss the 
dynamics and phases of the gauge theory and the physics of black holes \cite{Aharony:1999ti} 
-\cite{Hollowood:2006xb}.

In a recent paper \cite{Alvarez-Gaume:2006jg} two of us (PB, SRW) discussed the small Schwarzschild blackhole-string transition in 
$AdS_5\times S^5$ by relating it in the gauge theory to the large N \cite{GW, Wadia} transition, and its smoothening 
by non-perturbative effects. An important feature of studying the Schwarzschild blackhole in $AdS_5\times S^5$ is 
that the blackhole is uniformly spread on the $S^3$ at the boundary where the gauge theory lives. This is consistent 
with the fact that the scale invariance of the dual gauge theory is broken only by thermal boundary conditions. 
Consequently one can discuss the blackhole-string transition in terms of a unitary matrix model, where the unitary 
matrix (the thermal order parameter) is uniformly spread over $S^{3}$. 

In this work we would like to focus on confining gauge theories which have two length scales: the 
confinement scale $\Lambda$ and the temperature $T=\beta ^{-1}$. The relevant order parameters in such theories 
have a spatial variation on the scale of $\Lambda^{-1}$. The gravity duals of these theories have blackhole solutions 
which are localized on the boundary. 
It has been argued in \cite{Nastase:2005rp, Aharony:2005bm}, that 
their holographic dual corresponds, in the large N limit, to a localised region of 
the deconfinement phase. This object has been called the plasma ball in 
\cite{Aharony:2005bm}, and it has a mass and a lifetime of $o(N^2)$. 
A qualitative gauge theory discussion in \cite{Aharony:2005bm} uses a balancing of 
positive surface tension and negative pressure inside the plasma ball, 
to argue for its existence. 

There is no doubt that it is important to study the plasma ball and its dynamics. 
Besides its utility for the physics of gauge theories at finite temperature, 
it is one more concrete laboratory for testing and studying various conundrums 
presented by blackholes \cite{David:2002wn}.  The fact that the blackhole dual is localized on the boundary provides 
a greater handle on studying the horizon and what lies behind it.

Before we begin to make headway into an understanding of these problems, 
we need to have a dynamical handle on the plasma ball in the gauge theory. 
This is a standard hard strong coupling problem. Here by strong coupling we mean 
large t' Hooft coupling $\lambda = Ng_{YM}^2$. One natural strategy is to use numerical techniques. 
However a direct numerical approach is also difficult without developing a 
formalism within which we can ask the right questions. 

We present a partial answer to 
this question in this work. We will discuss the plasma ball as a 
{\it large N soliton} which can be discussed in terms of various order parameters which 
distinguish between the confinement or deconfinement phases of the gauge theory. 
In order to do a concrete calculation we will focus on a concrete model 
that was discussed in \cite{Aharony:2005bm} in which an interpolating solution was found 
between two bulk solutions of type IIB string theory: the AdS soliton \cite{Horowitz:1998ha} and a blackbrane. 
Both solutions are asymptotically $R^2\times S_{\tau}^1\times S_{\theta}^1$, 
where $S_{\tau}^1$ is the thermal circle of radius $\beta$ and $S_{\theta}^1$ is a 
Scherk-Schwarz spatial circle of radius $2\pi$. The corresponding gauge theory is a 
Scherk-Schwarz compactification of ${\cal N}=4$ $SU(N)$ gauge theory, 
on $R^2\times S_{\tau}^1\times S_{\theta}^1$. The relevant and natural order 
parameters of this gauge theory are the holonomies of the gauge field around 
$S_{\tau}^1\times S_{\theta}^1$. In fact for technical reasons we will compactify 
$R^2$ to a Scherk-Schwarz cylinder, so that the Euclidean spacetime of the 
gauge theory is $R^1\times S_{\tau}^1\times S_{\theta}^1\times S_{\alpha}^1$. 
The radius of $S_{\alpha}^1$ is chosen larger than that of the $S_{\tau}^1$ and $S_{\theta}^1$
\footnote{This additional compactification of the boundary does not disturb the bulk solution}.

We discuss the effective action of the gauge theory in the long wavelength expansion 
defined by the confinement scale $\Lambda$. The effective action, in the axial gauge along 
the non-compact direction $x$, is a one dimensional model of three unitary 
matrices $U$($x$), $V$($x$) and $W$($x$) corresponding to the zero modes of the Wilson loops on 
$S_{\tau}^1\times S_{\theta}^1\times S_{\alpha}^1$. Using the fact that we are working 
with a confining gauge theory of adjoint fields which are all short ranged (of the order of $\Lambda^{-1}$) one can integrate out V($x$) and W($x$) to 
arrive at an effective action involving the single unitary matrix U($x$), which has the general form 
\begin{equation}
S = \Lambda^{-1}\int^{\infty}_{-\infty} dx f(U) \tr(\partial_{x}U\partial_{x}U^{\dagger}) + g (U)
\label{effaction}
\end{equation}
where $\Lambda^{-1}$ is the confinement scale, and $f(U)$ and $g(U)$ are  
gauge invariant functions of $U$. $f(U)$ and $g(U)$ contain the information that the gauge theory has a first order confinement/deconfinement phase transition. 

It is possible to discuss soliton solutions of the
general multi-trace model using the Hamiltonian formulation together with the 
method of dealing with multi-trace operators developed in \cite{Alvarez-Gaume:2006jg}. 
However in order to exhibit a solution we simplify the effective action even further and 
present the soliton (plasma kink) solution. It turns out to be the motion of 
the Fermi surface of the many fermion problem that is equivalent to the matrix model 
in the $SU(N)$ invariant sector. This solution interpolates between the 
confinement and deconfinement phases and has energy density peaked at the phase boundary. 

In our investigations we realized that it is imperative to use the $2+1$ dimensional phase space 
formulation of the classical Fermi fluid theory. The collective field formalism, which is a hydrodynamical 
description in $1+1$ dimensions inevitably leads to shock formation and singularities. 
It is not clear whether a finite energy density soliton solution can be obtained within collective field theory. 
The shocks are spurious singularities due to the collective field description which 
correspond to the folds on the Fermi surface, which are inevitable.

The plan of the paper is as follows. In section \ref{plasma} we describe the
two bulk geometries- the AdS soliton \cite{Horowitz:1998ha} and the black brane solution, 
for which an interpolating domain wall solution 
was constructed in \cite{Aharony:2005bm}. In section \ref{eff}, we present a qualitative discussion as to how one 
can arrive at an effective description of the thermal gauge theory in terms of the holonomy matrices around the 
various cycles of the boundary, starting from a four dimensional gauge theory 
compactified on Scherk-Schwarz circles. For technical reasons we will be working with a 
gauge theory compactified on two Scherk-Schwarz circles. One can have two dual 
effective descriptions, in terms of either the Polyakov line or the Wilson loop over the spatial cycle. 
We present the general class of such effective matrix models. In the following sections we will be working 
with a particular matrix model belonging to this class. This model can be discussed in terms of an 
exact fermionic description
\cite{Brezin:1977sv, Wadia, Sengupta:1990bt, Gross:1990st, Polchinski:1991uq, Dhar:1992rs,  
dhar}. We shall also discuss the collective field equations \cite{Jevicki:1979mb} and indicate that their solution 
develops shocks in finite time. 

In section \ref{phasesmodel} we discuss the phase structure of the model. This model has two stable phases: the 
confined and deconfined phases, and it undergoes a first order phase transition at a particular 
temperature. Later in the section  we construct the soliton (kink) solution which interpolates 
between the two phases at the phase transition temperature. We then discuss some of the 
properties of the solution, in particular the surface tension
 of the soliton is discussed. We also present the localised (in one dimension) 
soliton solution at temperatures greater than the phase transition temperature, 
which approaches the confined phase in the two ends, and discuss some of it's properties. 

In appendix \ref{appen:expro} we show that starting 
from the confined phase of the theory, where the density of eigenvalues of the Polyakov 
line is uniform, we reach the clumped eigenvalue distribution only asymptotically, 
and never in any finite time. In appendix \ref{appen:collec} we discuss the relation of the shocks formed 
in the collective field theory description to the formation of folds in the Fermi description. 

\section{Plasma ball in the large $N$ gauge theory and dual black holes}\label{plasma}

A plasma ball is a localized spherically symmetric bubble of the deconfining 
phase of a confining 
gauge theory. In \cite{Aharony:2005bm} using the AdS/CFT correspondence, their existence was 
inferred by exhibiting a bulk solution that interpolates between the AdS 
Soliton\cite{Horowitz:1998ha} and the black-brane solution. 
The AdS soliton (AdSS), is given by the metric,
\begin{equation}\label{adss}
ds^{2} = L^{2}\alpha^{'}(e^{+2u}(d\tau^{2}
+T_{2\pi}d\theta^{2}+d\omega^{2}_{i}) +\frac{1}{T_{2\pi}(u)}du^{2})
\end{equation}
where,
\begin{equation}
T_{2\pi}(u)=1-(\frac{1}{2}(d+1)e^{u})^{-(d+1)}
\end{equation}
In this paper we will be working with $d= 3$. The coordinate $\theta$ 
is periodic with periodicity $2\pi$, and $\tau$ is the angular coordinate along the thermal circle of the Euclidean 
theory, with periodicity $\tau\rightarrow \tau +\beta$, and the $\omega_{i}$ are
the two non-compact coordinates, while $u$ is the radial coordinate.
The boundary topology is $R^{2}\times S^{1}_{\tau}\times S^{1}_{\theta}$, where 
$S^{1}_{\tau}$ and $S^{1}_{\theta}$ are the thermal and spatial cycles respectively. 
From the expression for $T_{2\pi}$, one sees that the spatial circle shrinks to zero size 
at a finite value of $u$.

The black-brane (BB) geometry is given by the metric
\begin{equation}\label{bb}
ds^{2} = L^{2}\alpha^{'}(e^{+2u}(T_{\beta}d\tau^{2}
+d\theta^{2}+d\omega^{2}_{i}) +\frac{1}{T_{\beta}(u)}du^{2}
\end{equation}
with $T_{\beta}(u)=1-(\frac{\beta}{4\pi}(d+1)e^{u})^{-(d+1)}$. This metric continued to Lorentzian signature has a 
horizon. Notice that when $\beta =2\pi$, the two metrics \ref{adss} and \ref{bb} are 
simply obtained from one other by interchanging the thermal circle with the spatial circle. 
Since geometrically there is no difference between the two, 
the free energy of the two configurations must be the same at this temperature. 
For $\beta < 2\pi $, the free energy of the BB geometry dominates the path integral while 
for $\beta > 2\pi$, the free energy of the AdSS geometry is dominant. 
In \cite{Aharony:2005bm} a domain wall solution which interpolates 
between these two solutions was constructed. Clearly 
such a domain wall solution exists only for 
$\beta=2\pi$ when the free energy of the two phases is equal. 
The domain wall is independent of one of the non-compact direction and in the other 
non-compact direction the BB and AdSS geometry are asymptotically 
reached at the two ends. 

These solutions can be incorporated within the IIB string theory by compactifying on $S^{5}$, with the five-form 
RR flux turned on. This would then have a dual boundary description in terms of the 
Scherk-Schwartz compactification of the ${\cal N}=4$ $SU(N)$ SYM theory on a spatial 
cycle, with thermal boundary condition on both the
cycle $S^{1}_{\tau}$ and  $S^{1}_{\theta}$. The gauge theory lives on $R^{2}\times S^{1}_{\tau}\times 
S^{1}_{\theta}$. At 
$\beta=2\pi$ clearly the two circles are identical and can be interchanged.  

The above discussion suggests that a ball of large but finite radius of the 
deconfined plasma can occur as a solution to the finite temperature 
effective action of the gauge theory, at a temperature slightly above $T_{c}$. 
At $T=T_c$ there exists a kink solution interpolating between the confined and deconfined 
phases. 

\section{Gauge theories on $R^{2}\times S^{1}_{\tau}\times S^{1}_{\theta}$}\label{eff}

From the AdS/CFT correspondence, these bulk geometries- the AdSS geometry and the BB geometry 
correspond in the thermal gauge theory to the confinement and deconfinement phases 
respectively \cite{Witten:1998zw}. These phases are characterised by the expectation 
value of the Polyakov line, which is the trace of the holonomy around the thermal circle,

\bea
U(w_{1},w_{2},\theta) =  {\cal P}\exp(-\oint A_{\tau} d\tau)
\eea
$w_{i}$ are the two non-compact coordinates and the $\theta$ is the angular coordinate along the spatial 
circle, while ${\cal P}$ denotes path ordering.  In 
particular, the expectation value of $\tr\ U$ vanishes in the confined phase while in the deconfined phase it takes a 
non-zero value \footnote {This 
basically reflects the fact that a quark in the fundamental representation of SU(N) has infinite free energy in the confining phase and finite free energy in 
the deconfined phase}.  
 Similarly one can define the holonomy around the spatial cycle $S^{1}_{\theta}$.
\bea
V(w_{1},w_{2},\tau) =  {\cal P}\exp(-\oint A_{\theta}d\theta)
\eea
Since the role of the two circles are interchanged in the two bulk geometries, it follows from the AdS/CFT 
correspondence, that $\tr V=0$ in the deconfined phase, and it is non-zero in the confined phase \footnote{This 
reflects a gluon condensate in the vacuum of the gauge theory \cite{Saviddy}.}. 
At $\beta=2\pi$, because the two geometries are identical under the 
interchange of the thermal and spatial circles, the effective action in terms of $V$ should 
be identical to the one in terms of $U$.
Later in this section we will qualitatively argue as to how one can arrive at an 
effective action in terms of both $U$ and $V$ and then in terms of either $U$ or $V$, starting from the 
four-dimensional gauge theory.  

Since we will mainly be interested in the solution which interpolates 
between the confinement and deconfinement phases as a function of one of the non-compact 
direction, 
it should be possible to find the one dimensional kink solution in an effective one-dimensional 
unitary matrix model. In order to realize this in a gauge theory at large $N$, it turns out to be 
convenient to work with $R\times S_{\tau}^{1}\times S_{\theta}^{1}\times 
S^{1}_{\alpha}$, where the $S^{1}_{\alpha}$ is the spatial circle, obtained by compactifying a noncompact direction 
previously labelled by the coordinate $w_{2}$. We introduce 
the holonomy along the spatial cycle $S^{1}_{\alpha}$ 
\bea
W(w_{1},\tau,\theta) = {\cal P}\exp(-\oint A_{\alpha}d\alpha)
\eea
This would correspond to replacing one of the non-compact directions of the bulk geometry that we 
discussed earlier, with a circle without changing the solution. Henceforth we shall set the noncompact direction 
$w_{1}\equiv x$.

\subsection{Effective action in terms of the Polyakov lines and Wilson loops}

The bosonic part of the action of the  general gauge theory will be written in terms of the gauge 
degrees of freedom 
$A_{1}$, $A_{\tau}$, $A_{\theta}$, $A_{\alpha}$ as well as the scalar fields 
$\Phi_{i}$ which transform 
in the adjoint representation. Here $A_{1}$ corresponds to the gauge field in the 
non-compact direction and we can choose the axial gauge $A_{1}=0$. These fields are in general functions of 
$(x,\theta,\tau,\alpha)$. Since the Scherk-Schwarz compactification breaks supersymmetry, 
the fermions are massive and the scalar fields get mass at one loop from quantum corrections. 
They can therefore be integrated out from the quantum effective action. 
Fourier expanding the gauge fields in all the circles and integrating out all the higher 
KK modes around every circle, we get a effective theory in terms of the zero modes of the 
fields: $A^{0}_{\tau}(x)$, $A^{0}_{\theta}(x)$, $A^{0}_{\alpha}(x)$.  

This effective theory in terms of the zero modes is gauge invariant, and therefore we should be able to write 
it down in terms of the zero modes of the the holonomy matrices $U$,$V$ and $W$. From now on we will 
use the notation $U$, $V$, $W$, to denote the zero modes of the above holonomy matrices.

The effective action will be a function of all possible gauge invariant operators. 
The gauge invariant operators are constructed out of the $Z_{N}$ invariant products of the 
polynomials of $U$, $V$ and $W$ and their covariant derivatives, $D_{x}U$, $D_{x}V$, $D_{x}W$, and are of the form
$\Pi_{i}\tr(U^{l_{i}} V^{m_{i}} W^{p_{i}} (D_{x}U)^{n_{i}}...)$, where  the exponents $l_{i}$, $m_{i}$, $p_{i}$, 
$n_{i}$,etc are integers, such that the sum of all the exponents $\sum_{i} l_{i}+m_{i}+p_{i}+n_{i}+... =0$. In the 
gauge $A_1=0$, the covariant derivatives are the same as the ordinary 
derivatives. 
At sufficiently long wavelengths we neglect the higher 
derivative terms which are suppressed by powers of the confining scale $\Lambda^{-1}$. 

Depending on which of the holonomy matrices condense, there will be three phases in the 
gauge theory. In the BB phase $\tr U\neq 0$, $\tr V=0$ and $\tr W=0$. In the other two phases one of the spatial 
holonomy matrices $V$ or $W$ will get 
expectation values, while the expectation value for the other two vanish.
However we are interested in an 
interpolating solution between the black brane and the AdS soliton,
and not in the transitions involving all the three cycles.
If we choose the radius $R_(S^{1}_{\alpha})>R(S^{1}_{\tau}),\ R(S^{1}_\theta)$, at the temperature of interest,  
then from the supergravity solution it follows that the cycle $S^{1}_{\alpha}$ never
shrinks and corresponds to $\vev {W}=0$ in the gauge theory. 
We can therefore put $W=0$ in the effective three matrix model to once again obtain a two matrix 
model. The action for this will in general be very complicated, with all terms that are 
allowed by gauge invariance. It will contain words of the type 
$\tr(U^{n_{1}}V^{n_{2}}U^{n_{3}}....)$, and also derivative terms. The general action in the long wavelength 
expansion will be of the form,
\bea\label{seff}
S_{eff}&=&\Lambda^{-1}\int dx f_{1}(U,V)\tr|\partial_{x}U|^{2}+f_{2}(U,V)\tr|\partial_{x}V|^{2}+ \\ 
\nn & & f_{3}(U,V)\tr(\partial_{x}U\partial_{x} V^{\d}) + f_{4}(U,V)+\textrm{h.c}
\eea
where, $\Lambda^{-1}$ is the confinement length scale, $f_{i}$'s are gauge invariant functions of arbitrary 
polynomials of $U$, $V$ and $\beta$ with appropriate factors of $N$. 
At $\beta=1/2\pi$, when the size of the two cycles are equal, the effective 
action will be invariant under $U \Rightarrow V$.  
Integrating over either the $U$ or the $V$ we will get a single matrix model in terms of 
$V$ or the $U$ matrix. 

 Since the theory is confining and has a mass gap, we can integrate out the $V$ matrix, without worrying about 
infrared divergences, and we will be left with a model, given by 
\begin{equation}
S = \Lambda^{-1}\int dx f(U) \tr(\partial_{x}U\partial_{x}U^{\dagger}) + g (U)
\label{effaction1}
\end{equation} 
where again $\Lambda^{-1}$ is the confinement scale, and $f(U)$ and $g(U)$ are temperature 
dependent, gauge invariant functions. As in 
(\ref{seff}) we have neglected all the higher derivative terms in the action, which are suppressed by powers 
of the confining scale. Equivalently we could integrate out $U$ to arrive at a matrix model of $V$.   
\par

 In the sequel we will mainly study the soliton solution of the simplest of this class of models, 
given by \footnote{This model has previously appeared in the discussion of 1+1 dimensional gauge theories 
\cite{Semenoff:1996xg}}.

\begin{equation}
S=\Lambda^{-1}\int dx N\tr(|\partial_{x}U|^{2}) + \xi|\tr U|^{2}
\label{zareq}
\end{equation}
Here we will assume that $\xi >0$ which ensures the existence of a first order 
phase transition at some value of $\xi$. By rescaling $x\rightarrow \Lambda^{-1} x$, 
we can remove the explicit  $\Lambda$ dependence from the above action to get,
\begin{equation}
S=\int dx N\tr(|\partial_{x}U|^{2}) + \xi|\tr U|^{2}
\label{zareq1}
\end{equation}
where $x$ is now given in units of $\Lambda^{-1}$. Hence forth we will be 
using this form of the action\footnote{Therefore all the quantities we calculate later in the 
text like the surface tension of the phase boundary of the soliton, for example, will be 
given in units of the confinement scale.}. 
\section{Analysis of the one dimensional matrix model}\label{phasesmodel}

In this section we will analyze the phase diagram of the unitary matrix model given by the action of the 
form (\ref{zareq}). 
The matrix model described by the action(\ref{zareq}) can be discussed using two methods. One is to use the 
collective field theory techniques as was done by Jevicki and Sakita\cite{Jevicki:1979mb}. 
This is basically a collective field description in $1+1$ dimension. The Hamiltonian is written in terms of the 
density $\rho(\theta,x)$ and velocity $v(\theta,x)= \partial_{\theta}\Pi(\theta,x)$, 
where $\Pi$ is the canonical conjugate of $\rho$. The $\rho(\theta,x)$ field is the 
eigenvalue density field constructed out of the matrix $U$,

\begin{equation}  
\rho(\theta,x) = \sum^{+\infty}_{n=-\infty} \rho_{n}(x) e^{2i\pi n\theta}
\end{equation}
where $\rho_{n}=\frac{1}{N}\tr(U^{n})$. For example, from the matrix model described by equation 
(\ref{zareq}), we get the following collective field Hamiltonian,
\begin{equation}
H_{cf}= \int d\theta\ (\frac{\rho v^{2}}{2} +\frac{\pi^{2}\rho^{3}}{6}) -\xi |\rho_{1}|^{2}
\label{ham}
\end{equation}

This Hamiltonian, gives rise to the following set of fluid dynamical equations,

\bea\label{bursou}
\frac{\del \rho(x,\theta)}{\del x}+ \frac{\del}{\del \theta}(\rho(x,\theta) v(x,\theta)) &=&0 \\
\frac{\del v(x,\theta)}{\del x}+ v(x,\theta) \frac{\del v(x,\theta)}{\del \theta}+\pi^2
\nn \rho(x,\theta) \frac{\del \rho(x,\theta)}{\del \theta} &=& -2 \xi \rho_1(x) \sin\theta
\eea
Here $\theta$ is a periodic variable defined in the range $[-\pi,\pi]$ and $x$ is 
a variable defined in the range 
$(-\infty,+\infty)$. The collective field approach is only 
valid for solutions which are spatially uniform,(for which $v(x,\theta)=0$ and
$\frac{\del}{\del x} \rho(x,\theta)=0 $). 
The spatially non-uniform solutions generically develop shocks in finite time, after which the collective field equations are not valid. \footnote{ As discussed in more detail in  appendix \ref{appen:collec}, this phenomenon can be understood from the underlying fermionic theory. Infact if we change $x \rightarrow ix$ and $v \rightarrow -iv$ in equation (\ref{bursou}), we get the inviscid Burgers equation with a source term.  In \cite{hodograph}, it has been shown, using the method of hodograph transformation, that the source free version of the Burgers equation develops shock in finite time. }

A correct (and exact) way to analyze the model (\ref{zareq}) is to rewrite the model as a 
theory of interacting fermions \cite{Polchinski:1991uq} with the Hamiltonian (where the '$x$' direction is identified 
with the Euclidean time). 

\begin{equation}
H = \int d\theta \psi^{\dag}(\theta)\partial^{2}_{\theta}\psi(\theta) 
-  \xi |\int d\theta e^{i \theta} \psi(\theta) \psi^{\dag}(\theta)|^2 
\end{equation}

In the large $N$ limit the fermion system will be classical and 
one can use the phase space density, ${\cal U}(p,\theta,x)$ such that,
\bea
\int \frac{dp}{2\pi} d\theta \ {\cal U}(p,\theta,x)= 1
\label{norm}
\eea
 If a phase space cell is occupied then ${\cal U}(p,\theta,x)=1$ or else ${\cal U}(p,\theta,x)=0$. Hence 
${\cal U}(p,\theta, x)$ satisfies the relation\footnote {The relation (\ref{U}) is true only 
at large $N$. At finite $N$, ${\cal U}$ satisfies the relation, ${\cal U}*{\cal U}\equiv \cos\frac{1}{2N}
(\partial_{\theta}\partial_{p'}-\partial_{p}\partial_{\theta'})[{\cal U}(p,\theta){\cal U}(p',\theta')]
|_{p'=p,\theta'=\theta}={\cal U}$ \cite{dhar}, which reduces to equation (\ref{U}) at large $N$.}
\bea
{\cal U}(p,\theta,x)^2={\cal U}(p,\theta,x)
\label{U}
\eea
The Hamiltonian written in terms of the phase space density is,
\begin{equation}\label{hamil}
\frac{H}{N^2}= \int dp d\theta \frac{p^{2}}{2} {\cal U}(p,\theta,x) - \xi | \int dp d\theta e^{i\theta} {\cal U}(p,\theta,x)|^2
\end{equation}
In terms of ${\cal U}(p,\theta,x)$, the density and velocity $\rho(\theta,x)$ and 
$v(\theta, x)$ are, 
\bea
\label{rho}
\rho(\theta,x)=\int \frac{dp}{2\pi} {\cal U}(p,\theta,x), \ v(\theta, x)=\frac{1}{\rho}\int 
\frac{dp}{2\pi} p\ {\cal U}(p,\theta,x)
\eea 

In the appendix we will further discuss the relation between the phase space and collective field theory 
approach and we will interpret the shock formation as the formation of folds on the Fermi surface. 
Hence the shock singularities are artifacts of the collective field approach and
are resolved by a more accurate treatment. 
\par 
In the following sections we will analyze the solutions of the fermionic Hamiltonian (\ref{hamil}). 
We will start by describing the spatially uniform solution (phases of the theory) and then describe the 
non-uniform interpolating solution (plasma kink). 
\par

\subsection{Spatially uniform solutions}\label{sus}
Here we analyze those solutions where the  density of the eigenvalues of the matrix $U$ is uniform over the direction $x$. 
In this case the location of the Fermi level will also be constant in $x$. Classically $\rho$ can always be chosen to be an even function of $\theta$. Then the potential in the equation (\ref{hamil}) becomes,
$\xi (\int  dp\frac{d\theta}{2\pi} \cos \theta {\cal U}(p,\theta,x) )^2$. In 
the Hatree Fock approximation, the phase space evolution equation 
for a single particle is, 
\bea
\dot{\theta}&=&p \\
\nn  \dot p &=& - 2 \xi \rho_1(x) \sin \theta
\eea
Where $\rho_1(x)=\int dp\frac{d\theta}{2\pi} \cos \theta {\cal U}(p,\theta,x)$ and 
$\dot{\theta}\equiv\frac{d}{dx}\theta$, $\dot{p}\equiv \frac{d}{dx}p$. For a spatially uniform solution, 
$\rho_1$ is independent of $x$ and we can integrate the above equations to get,
 \begin{equation}\label{eom}
p^{2}=2(E+ 2 \xi \rho_1 \cos\theta)
\end{equation}
where $E$ is the energy of the particle. Therefore for a particle on the Fermi level, we have,
\bea \label{fermi}
\hat{p}_{\pm}= \pm\sqrt{2(E_f+2\xi \rho_1 \cos\theta)}
\eea
where $\hat{p}_{\pm}$ correspond to the upper and lower branches of the Fermi level. 
Consequently 
\bea\label{static}
\rho(\theta)= \frac{\sqrt{2}}{\pi}\sqrt{E_f+2\xi\rho_{1}\cos\theta}. 
\eea
One has to satisfy the normalization condition given in equation (\ref{norm}) and the 
self consistency condition for $\rho_1$, which effectively solves $E_f$ in terms of $\xi$ and $\rho_{1}$
\bea \label{phases}
\int d\theta \frac{\sqrt{2}}{\pi}\sqrt{E_f+2\xi\rho_{1}\cos\theta}=1 \\
\nn \int d\theta \frac{\sqrt{2}}{\pi} \cos\theta \sqrt{E_f+ 2 \xi\rho_1 \cos\theta}=\rho_1
\eea
Depending on whether $|\frac{E_f}{2\xi\rho_{1}}|< 1$ or $|\frac{E_f}{2\xi\rho_{1}}|\ge 1$, 
the integrals in equation(\ref{phases}) will be 
evaluated between the limits $[-\theta_0,\theta_0]$, with $\theta_{0}< \pi$, or over the full range $[-\pi, +\pi]$. 
The former case corresponds to the gapped phase, as $\rho(\theta)=0$ outside$[-\theta_0,\theta_0]$.), while the 
latter case corresponds to the ungapped phase) \cite{Wadia}.
\par

One can study the different static phases of the model, by solving the self-consistency and the normalization conditions 
given in equation (\ref{phases}) simultaneously. This is hard to do analytically, but can be studied numerically. 
However it would be useful to have an understanding of the various phases as  extrema of the potential in terms of 
$\rho$.
This potential can be obtained from the Hamiltonian given in equation (\ref{hamil}), 
using equations (\ref{rho}, \ref{fermi}) to integrate over $p$. We then obtain,

\bea\label{haam}
{\cal H} =\int d\theta \frac{1}{2}\rho v^{2} + V([\rho])
\eea
where, 
\bea\label{pot}
V([\rho]) = \int d\theta \frac{\pi^{2}\rho^{3}}{6} -\xi |\rho_{1}|^{2}
\eea 

The potential of the model is actually a function of the infinitely many Fourier modes of $\rho$. 
Note that the static phases are all of the form given by the equation (\ref{static}). 
It is therefore useful to parametrize $\rho$ by 
\bea
\rho =\sqrt{\sum^{\infty}_{n=0} a_{n} \cos(n\theta)}
\eea
With this parametrization, the uniform phase solution is given by $a_{0}=\frac{1}{2\pi}$, 
and all other 
$a_{n}=0$, while the gapped phase corresponds to $a_{n}=0$, for $n>1$ and $a_{0}$, $a_{1}$ 
taking appropriate 
values. 
 With this parametrization, the potential will be a function of the $a_{n}$. 
Since all the phases of the theory lie 
in 
the plane given by $a_{n>1}=0$, it will be enough to restrict to this plane. 
We therefore parametrize 
$\hat{p}_{\pm}$ by 
 the following form.
\bea
\hat{p}_{\pm}= \pm\sqrt{2(E+2\xi C_{1}\cos\theta)}
\eea

\begin{figure}
\begin{center}
\includegraphics[height=4in,angle=270]{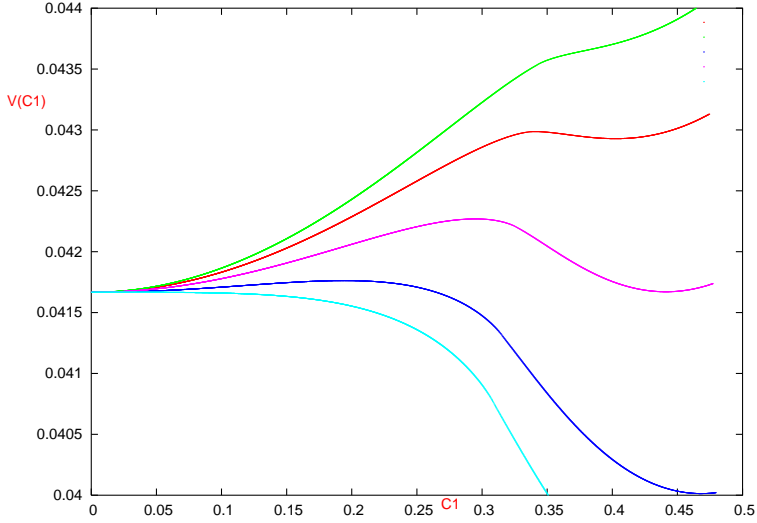}
\caption{Plot of $V(C_1)$ with $C_1$ with $\xi=0.22$, $\xi=0.23$, $\xi=0.237$, $\xi=0.245$ and $\xi=0.25$, with value 
of $\xi$ increasing from the top curve to the bottom.}
\label{fig:poten}
\end{center}
\end{figure}

 We determine $E$ in terms of $C_1$ by the normalization condition (\ref{norm}). 
Then substituting this in the expression for the potential, the potential becomes a 
function of only one parameter $C_{1}$. Then we can numerically calculate the 
potential given by the equation(\ref{pot}) as a function of $C_1$ (see figure \ref{fig:poten}). 

We now summarise the key points from our analysis of the phase structure of the model in consideration.
\begin{itemize}

\item{ At low enough values of $\xi$, there is a single phase where $\rho(\theta)=\frac{1}{2\pi}$ or 
$\hat{p}_{\pm}=\pm \frac{1}{2}$. Here $C_1=0$. 
This is the uniform phase of the eigenvalue distribution.}

\item {At $\xi=\xi_{n}=0.227$ there is nucleation of two phases for which $\rho(\theta)$ is no more a constant. 
Both the phases have a gapped eigenvalue distribution. One phase is 
unstable ($II$) and the other is stable ($III$).}

\item {The first order phase transition between the phase $I$ and phase $III$ occurs at 
$\xi=\xi_1=0.237$ and $C_1=0.4408$, $E=0.1711$.} 

\item {The phase $I$ becomes locally unstable at $\xi=\xi_{2}=.25$} 

\item{ At $\xi=\xi_{3}=0.23125$, and $C_{1}=0.3336$, phase $II$ has 
a gapped to ungapped transition, this is the point of the third order $GWW$ 
phase transition. \cite{GW, Wadia}}

\end{itemize}
\subsection{Spatially non-uniform solutions: plasma kinks}\label{plakin}

 In the previous section we have analyzed the phase structure of our model. In particular we saw that at 
$\xi=0.237$, the two stable phases (the confining and the deconfining phases) of the model have the same free energy. 
In this section  we will first describe an interpolating domain wall type solution 
from the deconfined phase to the confining phase, at this value of $\xi$. Later in the section 
we will also construct a localised soliton solution which reaches the confined phase for 
large values of $|x|$. 

\begin{figure}[h!]
\begin{center}
\includegraphics[height=1in]{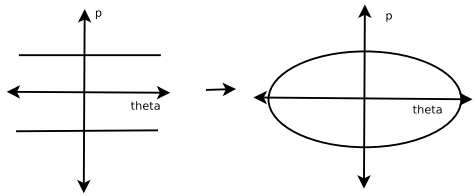}
\caption{A schematic picture of the Fermi levels.}
\label{fig:twofermi}
\end{center}
\end{figure}

The confining phase is described by a constant Fermi level which is given by the following  
equations in phase space,
\bea
\hat{p}_{\pm}=\pm \frac{1}{2}
\label{begin}
\eea
While in the deconfining phase, the Fermi levels were given by,  
\bea
 \hat{p}_{\pm}= \pm\sqrt{2(E+2\xi C_{1}\cos\theta)} 
\label{end}
\eea
Therefore we are looking for solutions in 
which the Fermi level evolves from (\ref{begin}) to (\ref{end}). In terms of the geometry of the Fermi level it is a 
evolution from a band like to an ellipsoidal structure fig. \ref{fig:twofermi}.

In terms of $\rho$, the solution has the property,
\bea
\rho(\theta, x) \rightarrow \frac{1}{2 \pi}, \quad x \rightarrow -\infty \\
\nn \rho(\theta, x)\rightarrow \frac{\sqrt{2}}{\pi}
\sqrt{E+2 \xi \rho_1 \cos\theta} , \quad x \rightarrow \infty
\eea 

Now in general the Fermi level will be described by the vanishing of some implicit function 
$f(\theta, p, x)=0$. In the static case, 
\bea
f(p,\theta) \equiv (p_{+} - \sqrt{E+2\xi C_{1}\cos \theta})
(p_{-} + \sqrt{E+2\xi C_{1}\cos \theta})=0
\eea
In the general case $f(p,\theta, x)$ is not of this simple form and may have more roots. 
This corresponds to the case where the upper and lower Fermi levels develop folds and become 
multi-valued in $\theta$. \footnote{In fact as is shown in appendix \ref{appen:collec} the 
folds are inevitably formed no matter what Fermi level configuration one starts with.} 

As each point in the phase space satisfies the equation, $\dot \theta=p, \dot p=V'(\theta)$, 
one can derive the time evolution of the function $f$ to be,   
\bea
\partial_{x} f +p\partial_{\theta}f + V'(\theta)\partial_{p}f =0
\label{evolution} 
\eea  
It would be interesting to try and solve the above equations numerically 
as a boundary value problem. We have not been able to do this. Instead we 
take a variational approach to the problem, and make a simple but reasonably accurate ansatz for the Fermi level. We will now summarise the main steps of the analysis.

\begin{itemize}

\item {We choose an ansatz for the Fermi level similar to the form in the static case,
\bea
f(p,\theta, x) \equiv (p_{+} -\rho(\theta, x) +v(\theta, x))(p_{-} + \rho(\theta, x) + 
v(\theta, x)) =0
\eea 
with $\rho$ given by, 
\bea
\rho = \frac{\sqrt{2}}{\pi}\sqrt{E(x)+2 \xi C_1(x) \cos\theta}
\eea
where the $E(x)$, $C_{1}(x)$ are functions of $x$. 
This would be a good approximation if the $E(x)$, $C_{1}(x)$ are slowly varying 
functions of $x$. 
What we are doing in effect is to approximate the actual solution by a two Fermi 
surface solution throughout the evolution of the system, always given by the two curves 
$p=\hat{p}_{\pm}$. Therefore $\hat{p}_{\pm}$ are of the form,
\bea
\hat{p}_{\pm}= \pm \sqrt{2}\sqrt{E(x)+2 \xi C_1(x) \cos\theta}+v(\theta,x) 
\label{ansatz}
\eea   
$E(x)$ is determined in terms of $C_1(x)$ by the condition (\ref{norm}) or equivalently
\bea
\nn \int d\theta \frac{\sqrt{2}}{\pi}\sqrt{E(x)+2 \xi C_1(x) \cos\theta} =1
\eea
We determine $v(\theta,x)$ by the continuity equation,
\bea
\frac{d}{dx} \int {\cal U}(p,\theta,x) dp \frac{d\theta}{2\pi} =0
\eea 
The solution of the continuity equation is given by,
\bea
v(\theta,x)=\frac{1}{\rho(\theta, x)}(\frac{\del}{\del x}\int_0^{\theta} d\theta \rho(\tilde{\theta}, x) 
d\tilde \theta) 
\eea
}

\item{Next, substituting this form of $\rho(\theta, x)$ and $v(\theta, x)$ back into the 
Hamiltonian and performing the $\theta$ integral, 
we get,
\bea
{\cal H}= {C'_1}^2 K(C_1)-V(C_1)
\label{hamilton}
\eea
where $C'_{1}=\frac{d}{dx} C_1(x)$. Hence the whole problem is reduced to a quantum mechanical 
problem of $C_1(x)$. The function $K(C_1)$ and $V(C_1)$ are determined numerically, 
and $K(C_{1})$ is positive and non-zero. Along the propagation in $x$ the quantity 
${\cal H}$ is conserved. This conservation law is used to determine the relation,
\bea 
\frac{d}{dx}C_1=\sqrt{\frac{E+V(C_1)}{K(C_1)}}
\eea
}
\item{The above equation is integrated numerically to obtain $C_1(x)$ as a function of $x$. 
Knowing $C_1(x)$ enables us to determine the phase space density ${\cal U}(p,\theta,x)$. 
The plot of $C_1(x)$ as a function of $x$ is shown in figure (\ref{fig:sol}). It should be 
noted that the soliton rises slowly but approaches the other end relatively fast. This follows 
from the asymmetric nature of the potential.   
}
\begin{figure}
\begin{center}
\includegraphics[height=4in,angle=270]{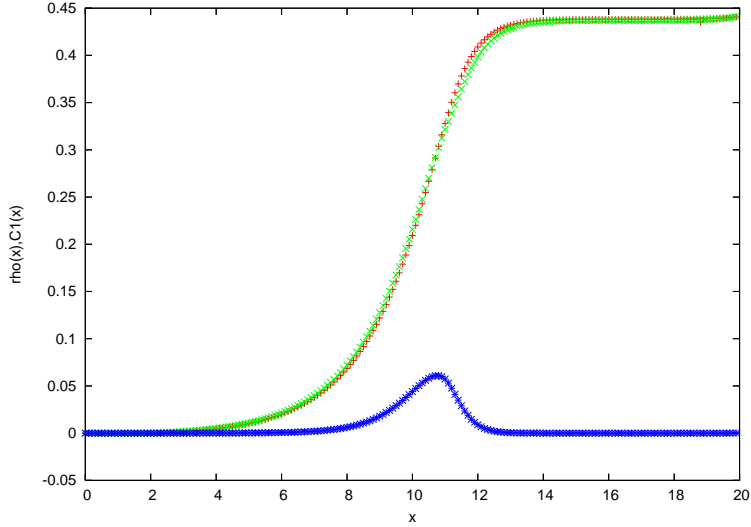}
\caption{Plot of $C_1(x)$ (green), $\rho(x)$ (red) and free energy density (blue, not in scale)}
\label{fig:sol}
\end{center}
\end{figure}

\item{It is important to check for the self consistency of this ansatz. This can be done by 
substituting the $\rho_{1}(x)$ obtained from our ansatz into the single 
particle equations and see how they evolve in $x$ under this $\rho_{1}$. 
One can then compute the $\rho_{1}(x)$ obtained from this exact evolution at each 
instance of $x$, which we denote by $\rho^{a}_{1}(x)$ and compare with $\rho_{1}(x)$ 
obtained from the ansatz. If $\rho_1(x)$ were an exact solution, then one would get 
$\rho^{a}_{1}(x)=\rho_{1}(x)$. This is checked numerically. 
We started with $50 \times 50$ particles uniformly distributed over the phase space 
region $p \in [-\frac{1}{2}, \frac{1}{2}], \theta \in [0,2\pi]$. 
This gives us the band like Fermi level in figure \ref{fig:twofermi}. We study the 
evolution of the individual particles under the driving force $2\xi\rho_1(x)$ and 
calculate the $\rho^{a}_{1}(x)$ from the phase space distribution of the particles. 
We present the plots comparing the two values of $\rho_{1}(x)$, in figure \ref{XXX}.}

\item{ One may also look at the snapshots of the phase space particles. In figures 
\ref{fig:phase} and 10 we have presented two snapshots taken at 
$x \approx 11.9$ and at $x \approx 11.6$. We see from the plots that the system is 
driven to the gapped phase configuration to a good accuracy. The  phase space 
snapshot at the later value of $x$ matches very well with the expected Fermi 
distribution in phase $III$ at $\xi=0.237$. This means that we are indeed 
reaching very near to the phase $III$. 

We also find that during the evolution of the Fermi sea, folds are formed on the 
Fermi level. As we discuss in the appendix \ref{appen:collec} this is inevitable. However 
the area under the folds is a small fraction of the area of the full Fermi surface. This 
shows that our ansatz of a Fermi level with no folds, is self-consistent. 

One also sees from the phase space plots that, as discussed in appendix \ref{appen:expro}, 
$\rho(0,x) \neq 0$ for all $x$.

\begin{figure}
\begin{center}
\includegraphics[height=4in,angle=270]{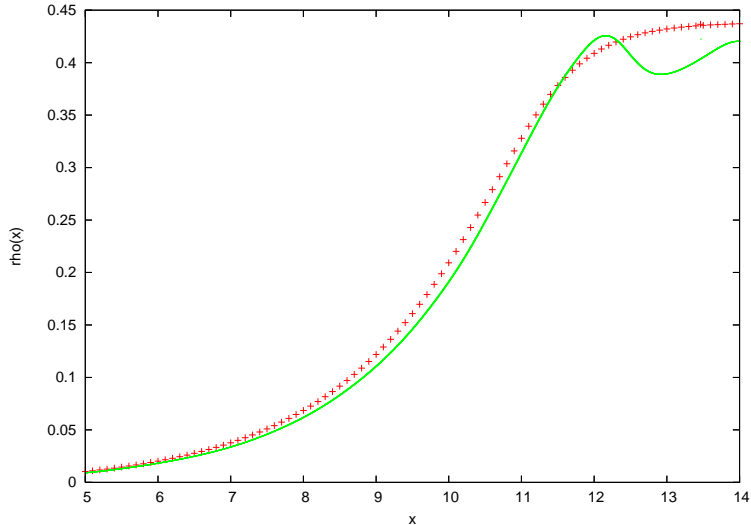}
\caption{Plot of $\rho^{a}_1(x)$ (green) and $\rho_1(x)$ (red) with $x$ }
\label{XXX}
\end{center}
\end{figure}
}

\item {If we continue to plot the evolution of the phase space particles for long times, we 
will see that the value of $\rho^{a}_{1}(x)$ will start falling from it's value in the gapped phase, 
and the particles will disperse away from the ellipsoid as the system will move away from the gapped phase. 
This happens because even though the $\rho_{1}$ we obtain from our ansatz drives the 
system very near to the gapped phase starting from the uniform phase 
(as is evident from the phase space plots), it does not take it exactly to the 
gapped phase, since no matter how good the ansatz is it is not the exact 
solution\footnote{This is clear since in the correct solution folds are 
always formed no matter how small.}. If we continue to the evolve the system 
this error will start accumulating and the system will again disperse away from the 
gapped phase. This 
problem would not occur if we could do the exact numerical simulation for the 
soliton in the phase space as a boundary value problem with value of $\rho^{a}_{1}(x)$ fixed at both ends.}
\end{itemize}

An important quantity that we can determine from our solution is the surface tension. 
The surface tension in general could either be positive or negative at the phase boundary. 
However, for lagrangians with positive kinetic terms, which is true in our case, the surface 
tension also turns out to be positive. 

In one dimension surface tension is defined as the total free energy of the soliton, 
which in turn is the total action for the soliton. 
Hence the surface tension $\sigma$ is, (see \cite{chaikin})

\bea
\sigma = 2 \int_{-\infty}^{+\infty} dx \left ( V(C_1(x))-V_{vacuum} \right) 
\eea 

This quantity at $\xi=\xi_1=0.237$ is numerically calculated to be, $\sigma=0.0027$.
\par

\subsection{Localized soliton- plasma ball }\label{plaball}
In the previous section we constructed an interpolating kink solution for $\xi =\xi_{1}$. For  $\xi$ between 
$\xi_1$ and $\xi_2$, the two minima corresponding to phase $I$ and phase $III$  have  different free energies 
(figure \ref{fig:poten}) and in particular, the minima corresponding to phase $I$($C_{1}=0$) is a false vacuum.  
In this case there exists a soliton solution which is localized in the $x$ direction, 
and which goes to $C_1=0$ at both $x\rightarrow \pm \infty$ \cite{raja}. 

\begin{figure}
\begin{center} 
\includegraphics[height=3in]{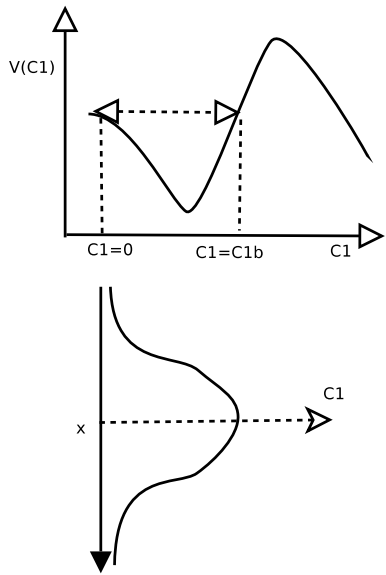}
\caption{Plot of $V(C_1)$ showing the bounce solution below.}
\label{fig:bsol}
\end{center}
\end{figure}

Such a solution has a simple interpretation in terms of a particle in real time moving in 
a potential $-V(x)$.
From the conservation of the Hamiltonian (\ref{hamilton}), it is obvious that if we start from $C_1=0$ at $x=0$, 
the solution never reaches phase $III$. It will bounces from a finite value of $C_{b}$ 
and comes back to the phase $I$ again, where $C_b$ is determined by the relation $V(C_b)=V(0)$. In Fig \ref{fig:bsol} we present a schematic plot of $-V(C_1)$ and the bounce solution. 

\begin{figure}
\begin{center}
\includegraphics[height=3in,angle=270]{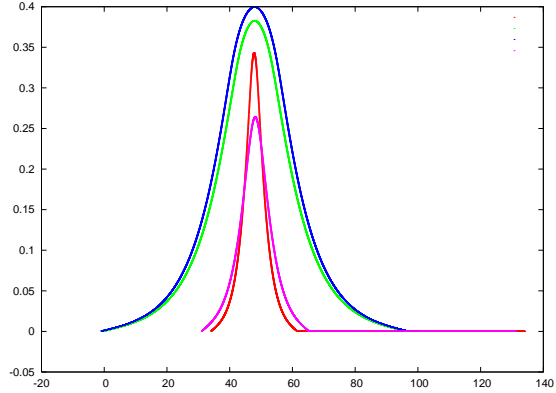}
\caption{Plot of $C_1(x)$ as a function of $x$ at $\xi=0.245$ (violet), $\xi=0.24$ (red), $\xi=0.238$ (green) and $\xi=0.2375$ (blue).}
\label{fig:losol}
\end{center}
\end{figure}

As before, one can construct such a solution numerically (see figure \ref{fig:losol}). 
This solution has a natural interpretation as a bubble of deconfined plasma within the 
confined phase. The plots shows two interesting trends. The first one is that the width and 
height of the soliton both increases as $\xi \rightarrow \xi_1 =0.237$ from above. 
The second one is that as $\xi\rightarrow \xi_2$, the height of the soliton decreases, but 
the width of the soliton also increases. Hence width of the soliton comes to a minimum at 
some value of $\xi$ between $\xi_1$ and $\xi_2$.

One can define the width $w$ of the localized soliton as a measure of the spread in $x$ over 
which the value of $C_1$ drops to a specified fraction $C_*$ of it's maximum $C_{b}$.
 As $\xi \rightarrow \xi_1$ the localized soliton becomes the semi-infinite soliton 
discussed in the previous section and consequently the width of the soliton goes to infinity. 
It would be interesting to calculate the change in the width of the soliton with 
$\xi$ as $\xi\rightarrow \xi_{1}$. In this limit,  $C_b$ almost reaches $C_{III}$. 
The equation of motion is given by,
\bea\label{local} 
2K(C_{1})C^{''}_{1}+ 2K'(C_{1})C^{'2}_{1} = V'(C_{1})
\eea
where $C'_1=\frac{d}{dx} C_1$, $K'(C_1)=\frac{d}{dC_{1}} K(C_1)$, and similarly 
$V'(C_1)=\frac{d}{dC_{1}} V(C_1)$. Expanding $K(C_1), V(C_1)$ around $C_1=C_{III}$, 
and using the fact that $V'(C_{III})=0$, and $C'_{1}$ will 
be small and negligible near $C_{1}=C_{III}$ (because $C_{III}$ is a turning point), we get  
from equation(\ref{local})
\bea 
\frac{d^{2}}{dx^{2}}\delta C_{1} = A(C_{III})\delta C_1
\eea                                       
where $\delta C_1 = C_{III}-C_1$ and $A= \frac{V''(C_{III})}{2K(C_{III})}$. 

Using the boundary conditions, $\delta C_1(0)= (C_{III}-C_b)$ and $\frac{d}{dx}{\delta C_1 }(0)=0$, 
one can solve the above equation to obtain,
\bea 
\delta C_1 =(C_{III}-C_b)(\cosh(\sqrt{A}x) 
\eea
If we define $B=C_b - C_{*}$, then the width $w$ is given by,
\bea
1+\frac{B}{C_{III}-C_b} =\cosh(\sqrt{A}w)
\eea 
Since $C_{III}-C_b\rightarrow 0 $ as $\xi\rightarrow \xi_1$, it follows that in this 
limit the leading $\xi$ dependence of $C_{III} - C_b$ will be of the form $C_{III}-C_{b}\sim (\xi-\xi_1)^a$, 
where $a$ could be any real positive number. Putting this dependence back into the above equation, 
and solving in the $w\rightarrow \infty$ limit, we get,
\bea
w \propto -\log (\xi-\xi_{1})                                                                                            
\eea
Hence we see that the width of the soliton diverges logarithmically with $\xi-\xi_{1}$.  

\section{Conclusion}\label{conc}

In this paper we have a presented a $o(N^2)$ soliton solution of a confining gauge theory which interpolates between 
the
confining and deconfinement phases separated by a first order phase transition. The soliton is a solution of the  
large
$N$, long wavelength effective action of the gauge theory expressed in terms of the thermal order parameter (Polyakov
line). The general three dimensional effective Lagrangian would have to contain higher derivative terms to support a 
soliton
solution and this would make the problem technically very difficult. However, in the present work we have analyzed a simpler one dimensional example. We have presented a qualitative discussion on the possible connection of this model with a higher dimensional confining gauge theory which has a gravity dual. The soliton that we have found numerically is a finite region of the deconfinement phase (plasma kink/ball)
with a positive surface tension at the phase boundary. The free energy density is also a smooth function every where in 
space.

Even though the soliton solution is obtained in a thermal gauge theory formulated in Euclidean spacetime it is 
reasonable
to expect it to be a static solution in Lorentzian spacetime at finite temperature.\footnote{In this case the 
holonomy
matrix $V(x)$ may be a more appropriate order parameter.} This fact can be inferred by observing that the bulk 
solution
can be analytically continued from Euclidean to Lorentzian spacetime. Given these facts it is tempting to identify 
the
phase boundary as dual to the horizon of the blackhole. A more precise understanding of this correspondence will 
enable
us to explore the structure of blackholes, especially `inside the horizon' and address very directly the persistent
question of the blackhole singularity.

\section{Acknowledgement}
We would like to thank Sriram Ramaswamy for useful discussion and guidance to the literature on the 
Burgers equation and the formation of shocks in fluid dynamics. We would like to acknowledge Avinash Dhar, 
Gautam Mandal and especially Shiraz Minwalla for very useful discussions. The research of SRW is supported 
in part by ``The J.C. Bose Fellowship'' of the Department of Science and Technology, Govt of India. 
PB would like to acknowledge CSIR for SPM fellowship.   

\appendix
\section{Analysis of the clumping in the eigenvalue distribution in finite time}\label{appen:expro}
In this appendix we will prove that if we give a small perturbation around phase $I$, $\rho(\theta,x)
$  
never becomes $0$ near the point $\theta=0$, at any finite $x$. 
Let us solve the equations of motion for individual phase space points near $\theta=0$. 
Near $\theta =0$ we can make the approximation, $\sin\theta \sim \theta$. 
The equations of motion can be written as,
\bea
\left( \begin{matrix} \dot p \\
                \dot \theta \end{matrix} \right) = M(x) \left( \begin{matrix}  p \\
\theta \end{matrix} \right)
\eea

where,
\bea
M(x)=
\left( \begin{matrix} 0 & 2 \xi \rho_1(x) \\
               1 & 0 \end{matrix} \right)
\eea
Here we start by approximating $\rho_1(x)$ with with a step function such that 
\bea
\rho_1(x)&=&\rho_1, \quad x>0 \\
         &=&0,\quad x<0
\eea
The solution of the equation is given by the condition,
\bea
\exp(- M x) \left( \begin{matrix}  p(x) \\
q(x) \end{matrix} \right)=\left( \begin{matrix}  p(0) \\
q(0) \end{matrix} \right)
\eea
If we look at the Fermi level given by, 
$\hat{p}_{\pm}(0)=\pm p_{0}$, then at "time" $x$ the position of the Fermi level will be,
\bea
\hat{p}_{\pm}(x)=\frac{\pm p_{0}}{\cosh(\sqrt{ 2 \xi x\rho_1 })}
\eea   

As $|\rho_{1}(x)| < 1$ , $\hat{p}_{\pm}(x)$ does not reach $0$ at any finite time. Similar result seems to be true 
for a 
time dependent $\rho_1$. Consequently, eigenvalue density function 
$\rho(\theta)=\hat{p}_{+}(\theta)-\hat{p}_{-}(\theta)$ 
is always non-zero at the point $\theta=0$. Hence any gap in the eigen value distribution can not open in finite time. However, the solution may asymptotically reach a gapped phase.

\section{Shock formation in the collective field equations and folds on the Fermi surface}\label{appen:collec}

In this section we will show that the collective field  equations develop shocks in finite time which can be understood from the underlying phase space picture as the formation of folds on the Fermi surface. The collective field equations may be derived from a classical theory of fermions. Consider first the theory of free fermions. We are looking at the phase space description of this theory. 
The motion of individual phase space points are described by the equations,

\begin{equation}
\dot{\theta} = p, \dot{p}=0
\end{equation}

From the above equation we can determine the equation of motion for a particle on the Fermi surface to be,   
\begin{equation}
\partial_{x}\hat{p} + p\partial_{\theta} \hat{p} = 0  
\end{equation}

where $\hat{p}$ denotes the value of $p$ at any point on the Fermi surface. 

\begin{figure}
\begin{center}
\includegraphics[height=2in]{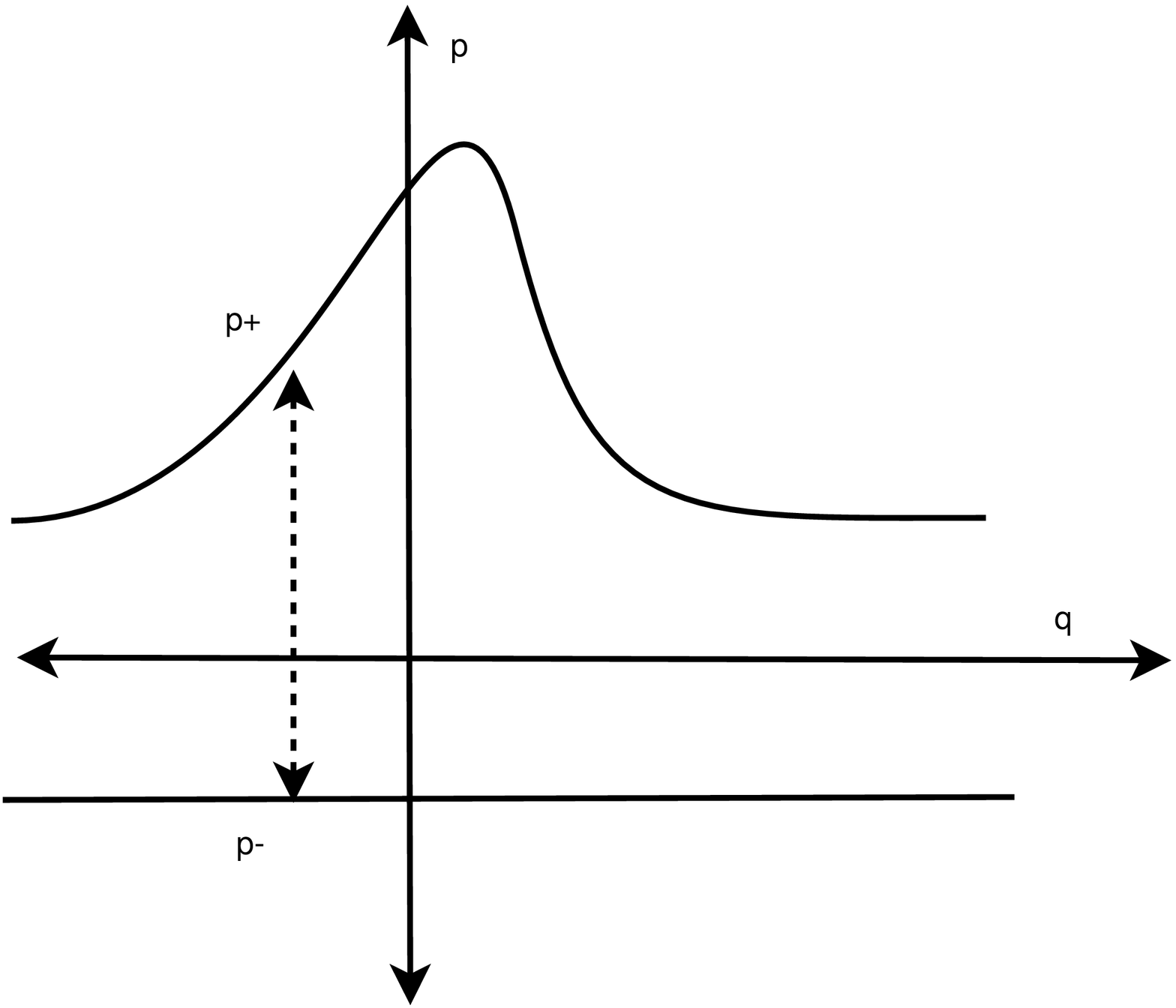}
\caption{Fermi level}
\label{fermi1}
\end{center}
\end{figure}

Now if the profile of the Fermi surface is such that for each value of $\theta$, 
there are exactly two points lying 
on the Fermi surface, one on the upper and lower Fermi level each (like in figure (\ref{fermi1})),
 then we have
\begin{equation}\label{hydro}
\partial_{x}\hat{p}_{\pm} + \hat{p}_{\pm}\partial_{\theta} \hat{p}_{\pm} = 0  
\end{equation}
where $\hat{p}_{\pm}$ characterize  the points on the upper and lower Fermi levels respectively. 
The source free version of the collective equations in (\ref{bursou}) are simply linear combination of the above two equations (see \cite{Polchinski:1991uq}), governing the dynamics of
 $\hat{p}_{+} + \hat{p}_{-}$ and  $\hat{p}_{+} - \hat{p}_{-}$, which are proportional to  
$v$ and $\rho$ respectively from (\ref{rho}). 

\begin{figure}

\begin{center}
\includegraphics[height=2in]{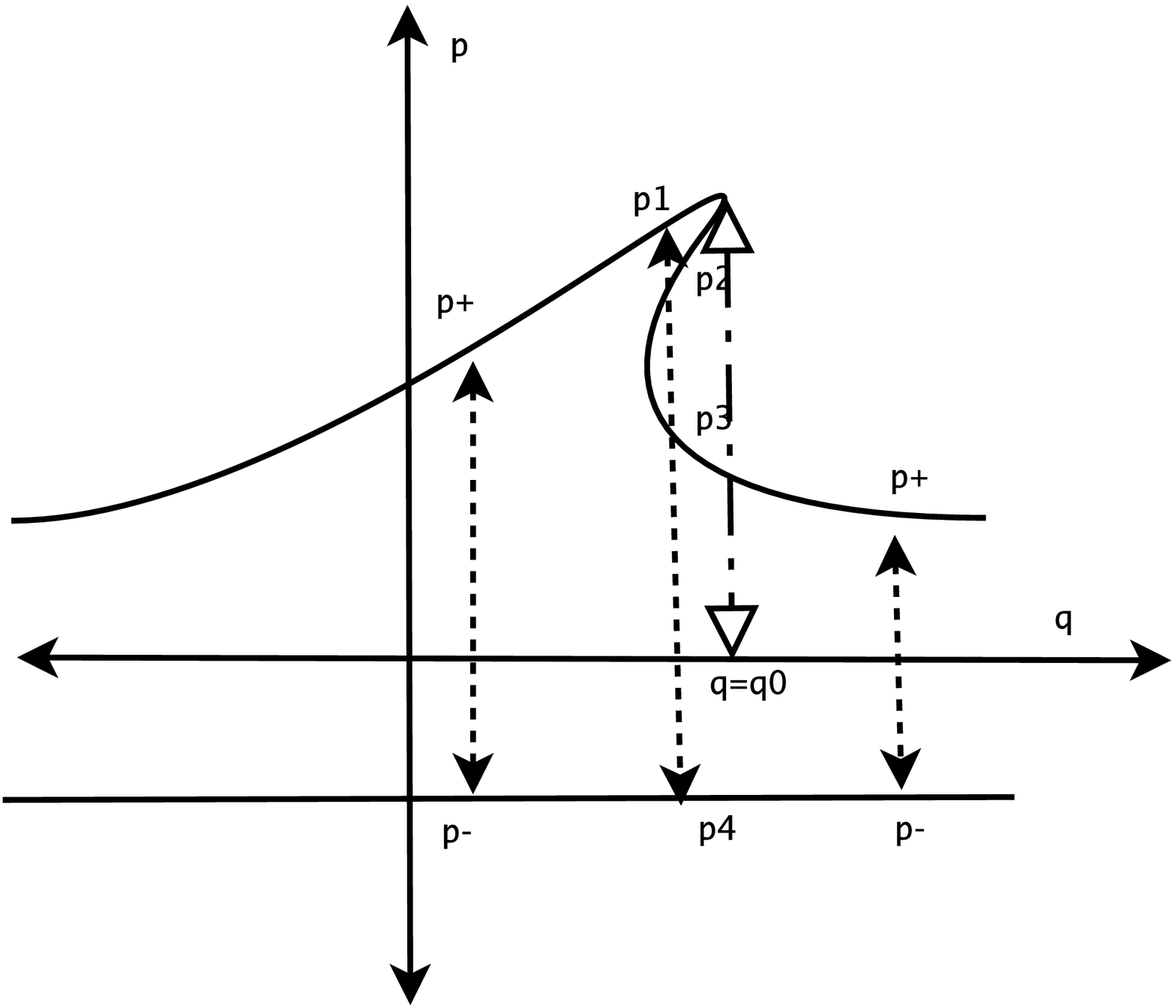}
\caption{Fermi level}
\label{fermi2}
\end{center}
\end{figure}
\par

This identification with the collective field equations is perfectly fine for a 
fluctuation of the form shown in the figure (\ref{fermi1}).  
However because of the equation of the motion, points of the curve which are higher, 
have greater velocity than the lower points, hence even if we start with a simple profile 
like that given in figure (\ref{fermi1}), the profile changes due to the unequal velocity of 
the various points lying on the Fermi level to a profile of the form given in figure (\ref{fermi2}). 
In figure(\ref{fermi2}), where the profile becomes multi-valued, the identification is not 
as 
before, since there are more than two values of $p$ corresponding to the same value of $x$. 
For instance, if at a point the Fermi profile has a multi valuedness of the "order four", 
that is there are four values of $\hat{p}$ corresponding to the same value of $\theta$, 
then the equation for $\rho$ becomes
\begin{equation}
2\pi\rho(\theta) = \int_{\hat{p}_{3}}^{\hat{p}_{4}} dp\  {\cal U}(p,\theta) +  
\int_{\hat{p}_{1}}^{\hat{p}_{2}} dp\  
{\cal U}(p,\theta)
\end{equation}
and similarly for the equation for $\rho v$. One can easily see that one cannot derive the 
simple collective field equations in this case. Hence the collective field equations do not describe the dynamics of the Fermi surface at all times.

However we can still look at the the topmost value of $p$ as $\hat{p}_{+}$ 
and the lowest value of $p$ as $\hat{p}_{-}$. In that case the equations governing the 
dynamics of 
$p_{+} + p_{-}$ and  $\hat{p}_{+} - \hat{p}_{-}$ are the same collective field equation 
throughout, but then we see clearly from figure (\ref{fermi2}). 
that the values of these variables jumps at $\theta=\theta_{0}$, 
and hence the $\theta$ derivative blows up at this point. This jump will correspond to the shock of the
collective field equations. 
Note that the description in terms of the fermion phase space is always perfectly smooth 
since it is after all the theory of free fermions.

In our case we are dealing with a $1+1$ dimensional interacting Euclidean fermionic 
theory 
given by a Lagrangian of one fermionic field $\Psi(\theta)$ 
\bea
{\cal L}=\int d\theta \Psi^{\dagger} \del_{x}\Psi+|\del_{\theta}\Psi|^2+2 \xi \int d\theta d\theta' 
\Psi^{\dagger}(\theta)\Psi(\theta)\cos(\theta-\theta') \Psi^{\dagger}(\theta')\Psi(\theta') 
\eea
These equations give rise to the equation of the form (\ref{bursou}). The phase space 
arguments discussed here will continue to hold even in this case again leading to shock formation in finite time (see fig$.10$). 
But the theory viewed as a theory of fermions will still be valid. 

\newpage
\begin{figure}[h!]
\begin{center}
\includegraphics[height=4in,angle=270]{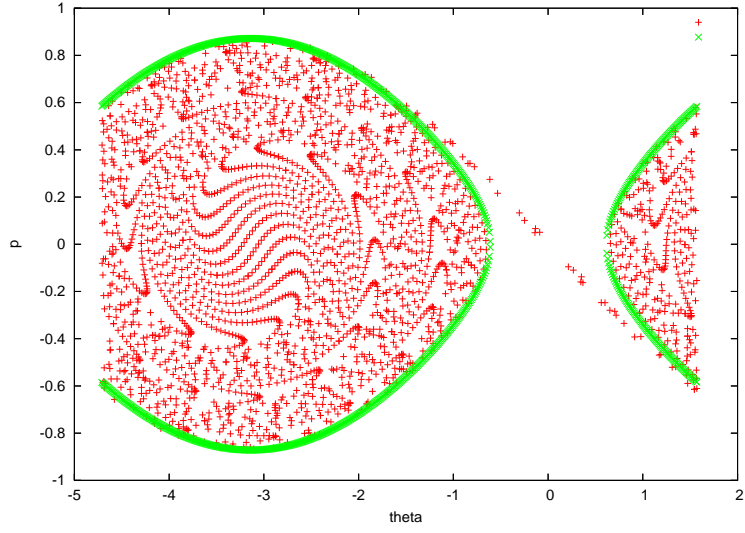}
\caption{Phase space particles (red) at $x \approx 11.9$ showing the match with Fermi surface (green) in phase III.}
\label{fig:phase}
\end{center}
\end{figure}

\begin{figure}[h!]
\begin{center}
\includegraphics[height=4in,angle=270]{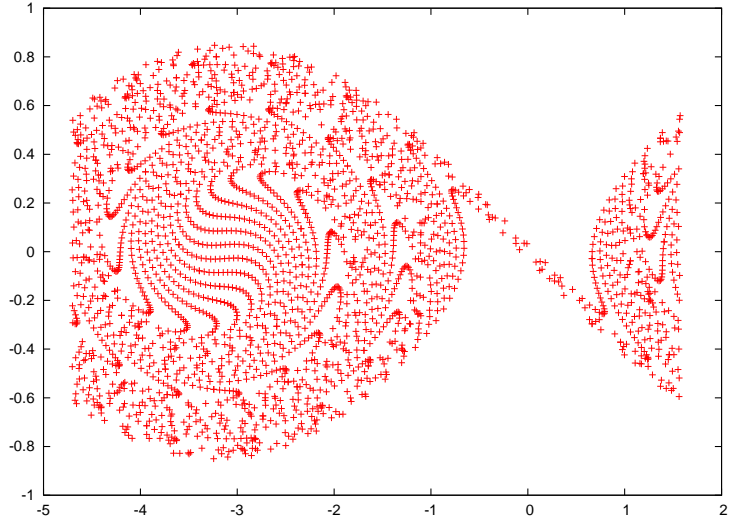}
\caption{Phase space particles at  $x \approx 11.6$ showing shocks at around $\theta \approx -0.8$ and $\theta \approx 0.8$.}
\end{center}
\label{fig:phaso}
\end{figure}

\end{document}